# Transmission resonances in silicon subwavelength grating slot waveguide with functional host material for sensing applications

S. Hadi Badri

Department of Electrical Engineering, Sarab Branch, Islamic Azad University, Sarab, Iran

sh.badri@iaut.ac.ir

## Abstract

A highly sensitive and selective $CO_2$ gas sensor is presented based on a subwavelength grating (SWG) slot waveguide. Polyhexamethylene biguanide (PHMB) as a functional material fills the slot gap as well as the space between the silicon pillars of the SWG structure. Beyond the photonic bandgap of the SWG slot waveguide, there are transmission resonances sensitive to the refractive index changes of PHMB due to the infiltration of $CO_2$ molecules into the functional material. The numerical simulations indicate that the sensitivity of the structure is $S$=12.9 pm/ppm which is considerably higher than the previously designed gas sensors based on functional materials. The higher sensitivity of the proposed sensor is attributed to the strong confinement of the light in the slot gap filled with functional material while previous designs have limited light-matter interaction by placing the functional material in the cladding. The proposed structure may be used to design various sensors by utilizing different functional material sensitive to the desired analyte.

## 1. Introduction

SWG in the silicon-on-insulator (SOI) platform consists of a periodic arrangement of high-index inclusion and low-index host materials with a period of $\Lambda$. This periodic arrangement of the two different materials provides an innovative tool for controlling light propagation and refractive index engineering. When the light, with an operating wavelength of $\lambda$, propagates along the periodicity of the longitudinal SWG, we can categorize the working regimes of the SWG waveguide as diffraction, Bragg reflection, and subwavelength regimes based on the given grating pitch ($\Lambda$) [1]. In the diffraction regime, the optical wave, with a wavelength shorter than the Bragg wavelength, radiates from the SWG structure into the surrounding environment. This effect has been exploited to implement optical fiber couplers [2, 3]. In the Bragg reflection regime, the periodic structure reflects the light due to the photonic bandgap effect. This regime has extensively utilized to design various elements such as lasers [4] and fiber Bragg gratings [5]. In the long-wavelength limit where $\lambda >> \Lambda$, the reflection and diffraction are negligible and the structure can be modeled with an equivalent homogeneous medium with an effective refractive index calculated by the effective medium theory (EMT) [1, 6]. The SWG structures in the subwavelength regime have been employed to implement gradient-index media [7-11].

Photonic integrated devices in the SOI platform rely on the light confinement in the silicon region due to its higher refractive index compared to the claddings. However, in some applications such as sensors, it is advantageous to change the guiding mechanism and confine the light in the low-index material [12, 13]. A slot waveguide has such a guiding mechanism and the light is strongly confined in the slot region. The slot waveguide is composed of two high-index rails, usually silicon, separated by a subwavelength gap filled with a low-index material. The rails and the slot region are surrounded by low-index claddings. The SWG slot waveguide is created by replacing the strip rails with the SWG structure. The SWG slot mode consists of surface-enhanced supermode, due to the high index-contrast leading to electric field discontinuity at the slot region, and Bloch mode due to the periodicity of the SWG structure [14]. The sensitivity of the output power of the SWG slot waveguide to the methane gas concentration has been investigated based on the slow-light effect by adjusting the SWG period to increase the group index [15]. In another study, the mode sensitivity of the SWG slot waveguide to the refractive index perturbations in the upper fluidic cladding has been analyzed [16]. Sensors based on other types of waveguides have also been studied. A ridge waveguide operating in the mid-infrared range overlapping with the characteristic absorptions of hydrocarbon gases, $C_2H_2$ and $CH_4$, has been presented [17]. A ridge waveguide has been modified into a dual hybrid plasmonic waveguide enhancing the evanescent field. This structure operates at 3.392 µm, an absorption line of methane gas, offering a sensitivity of 0.0715 mW/concentration [18]. Recently, different platforms for designing gas sensors have also been compared regarding their capabilities in creating a fully integrated spectroscopic setup [19].

In this paper, a highly sensitive sensor enabled by the functional material of PHMB and based on a silicon SWG slot waveguide operating in the Bragg reflection regime is presented. The slot gap of the waveguide is filled with PHMB which its refractive index decreases in the presence of the $CO_2$ gas. PHMB is also interlaced with the silicon pillars to increase the modal overlap with the sensing region. The previous designs of the $CO_2$ sensors have utilized the functional material of PHMB as the upper cladding limiting the interaction of the light and PHMB [20-22]. However, we exploit the strong light-matter interaction between the resonant transmission modes of the SWG structure and PHMB placed in the slot gap substantially increasing the sensitivity of the proposed sensor. The three-dimensional numerical simulations indicate that the sharp transmission peaks of the proposed structure are promising for sensing applications.

## 2. Subwavelength grating slot waveguide

To design transmission resonances based on the SWG slot waveguide for sensing applications, we choose the geometrical parameters of the SWG structure to operate in the Bragg reflection regime. Within the bandgap of the structure, the light is reflected back. However, beyond the bandgap, some transmission peaks appear due to the resonance of the light in the SWG structure. The schematics of an SWG slot waveguide in the SOI platform is illustrated in Fig. 1. The period of the SWG is $\Lambda$=415 nm operating in the Bragg reflection regime. The length of silicon pillars is $a=D\times\Lambda$=249 nm where the duty cycle is $D$=0.6. The thickness of the silicon rails and pillars is $h$=220 nm. A conventional slot waveguide is composed of two silicon rails with width of $w_{rail}$=400 nm separated by a slot gap of $w_{slot}$=100 nm placed on top of the silica substrate. To convert the mode of the conventional slot waveguide to the SWG slot waveguide, each rail is tapered to the SWG structure. The length of these tapers is $L_{taper}=10\times\Lambda$=4.15 µm while the length of the Bragg grating is $L_{BG}=28\times\Lambda$=11.62 µm. Although in the SWG slot waveguide, the light is largely confined in the slot region, however, the sensitivity of the structure increases as the modal overlap increases with the sensing region, i.e., the functional material. Therefore, as shown in Fig. 1, the functional material fills the slot as well as the spaces between the silicon pillars of the SWG claddings. PHMB, used as the functional material, exhibits reversible refractive index change in the presence and absence of $CO_2$ gas without the requirement of humidity. PHMB is a member of the guanidine polymer family that functions in both solution and solid states to reversibly bind with $CO_2$ gas molecules [21]. PHMB has two $CO_2$-binding amine groups in each monomer that binds $CO_2$ without catalysis by heat or humidity, and releases $CO_2$ when purged with $N_2$ or heated at low temperature. The refractive index of PHMB decreases as its density decreases due to the swelling of the polymer after the absorption of $CO_2$ gas molecules [23]. The resonance wavelength blueshift of gas sensors based on the functional material of PHMB is linearly correlated to the $CO_2$ concentration [20-22]. However, there has been no report of PHMB's sensitivity to other gases making the designed sensors based on PHMB highly selective to $CO_2$ gas.

The three-dimensional (3D) finite-difference time-domain (FDTD) method occupies less memory compared to the finite element method (FEM). Besides, the simulation time of the FDTD method is considerably shorter than the FEM method. Therefore, we used Lumerical FDTD software for simulating the structure. The FDTD simulations were performed using Lumerical's automatic non-uniform meshing algorithm with a minimum meshing step of 10 nm. The built-in material models of silicon and silica in Lumerical are used in simulations. We consider the refractive index of 1.55 at the wavelength of 1550 nm for PHMB in the absence of $CO_2$ gas [24]. A source mode is utilized to inject a transverse-electric (TE) mode to the slot waveguide. The main component of the TE mode is $E_x$. The coordinate axes are shown in Fig. 1. The perfectly matched boundary (PML) conditions are used to terminate the simulation domain. We should also emphasize that the wavelength step of 0.2 nm is used for transmission spectra calculations, however, only a handful of markers are used to represent the transmission curves. The transmission spectra of the SWG slot waveguides operating in the subwavelength and Bragg reflection regimes are compared in Fig. 2. When the grating pitch of the SWG structure is considerably lower than the operating wavelength ($\Lambda$=250 nm), the structure works in the subwavelength regime. In this case, the transmission of the SWG slot waveguide is higher than 0.91 in the wide bandwidth of 1300-1700 nm. However, when the structure operates in the Bragg reflection regime, the light is almost completely reflected in the wavelength range of 1388-1542 nm. For wavelengths shorter than 1388 nm, there are some ripples in the transmission spectrum, however, the transmission is lower than 0.134. In longer wavelengths ($\lambda$>1695 nm), the two waveguides with $\Lambda$=250 and 415 nm have similar performances. Beyond the photonic bandgap, the electromagnetic field at certain wavelengths forms a standing wave comprising of a forward and a backward propagating Bloch waves leading to transmission resonances [25]. It should be noted that in Fig. 2 the other geometrical parameters are the same for both structures, i.e., $a$=249 nm, $h$=220 nm, $w_{rail}$=400 nm, $w_{slot}$=100 nm, $L_{BG}=28\times\Lambda$ and $L_{taper}=10\times\Lambda$. In Fig. 3, the formation of the standing wave composed of a forward and a backward Bloch eigenmodes with large and nearly equal amplitudes for the first three transmission resonances are shown [26]. The electric field intensity of the first-order resonance mode at the wavelength of 1549.29 nm with the transmission peak of about 0.6 is shown in Fig. 3(a). The full-width at half-maximum (FWHM) of this transmission band is 2.96 nm. The second-order resonance mode occurs at the wavelength of 1563.8 nm with the transmission peak of 0.89 and FWHM of 5.51 nm as shown in Fig. 3(b). The third-order resonance mode at the wavelength of 1585.7 nm is also illustrated in Fig. 3(c). The electric field distribution and its components at the cross-section of the SWG slot waveguide are displayed in Fig. 3(d-g). It is obvious from this figure that the main component of the TE mode propagating through the SWG slot waveguide is $E_x$.

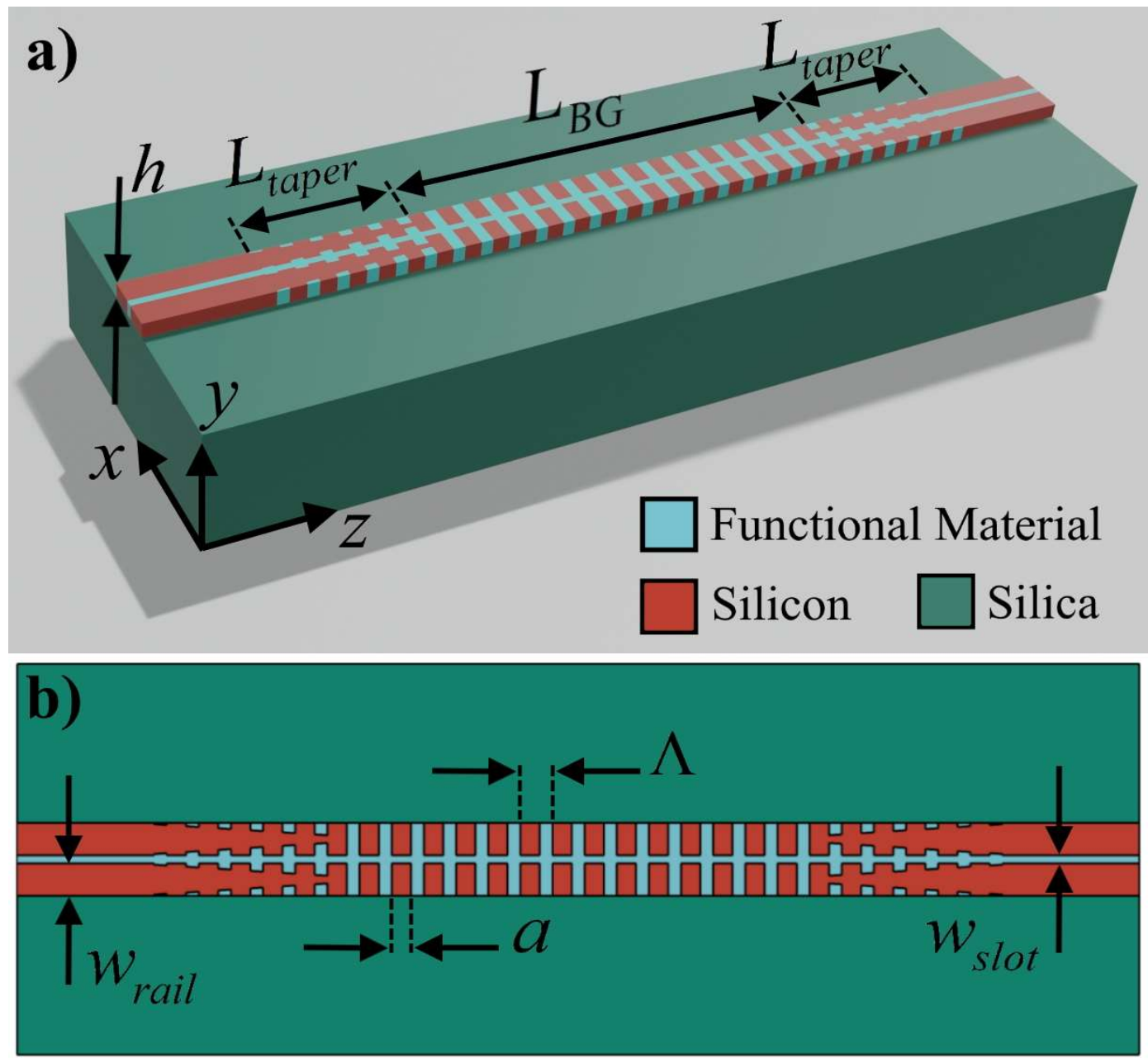

Fig. 1. a) Three-dimensional view and b) top view of the SWG slot waveguide. For illustration purposes, the length of the slot-to-SWG slot waveguide mode converter, $L_{taper}$, and the length of the Bragg grating, $L_{BG}$, are shorter in this figure compared to the structure used for simulations.

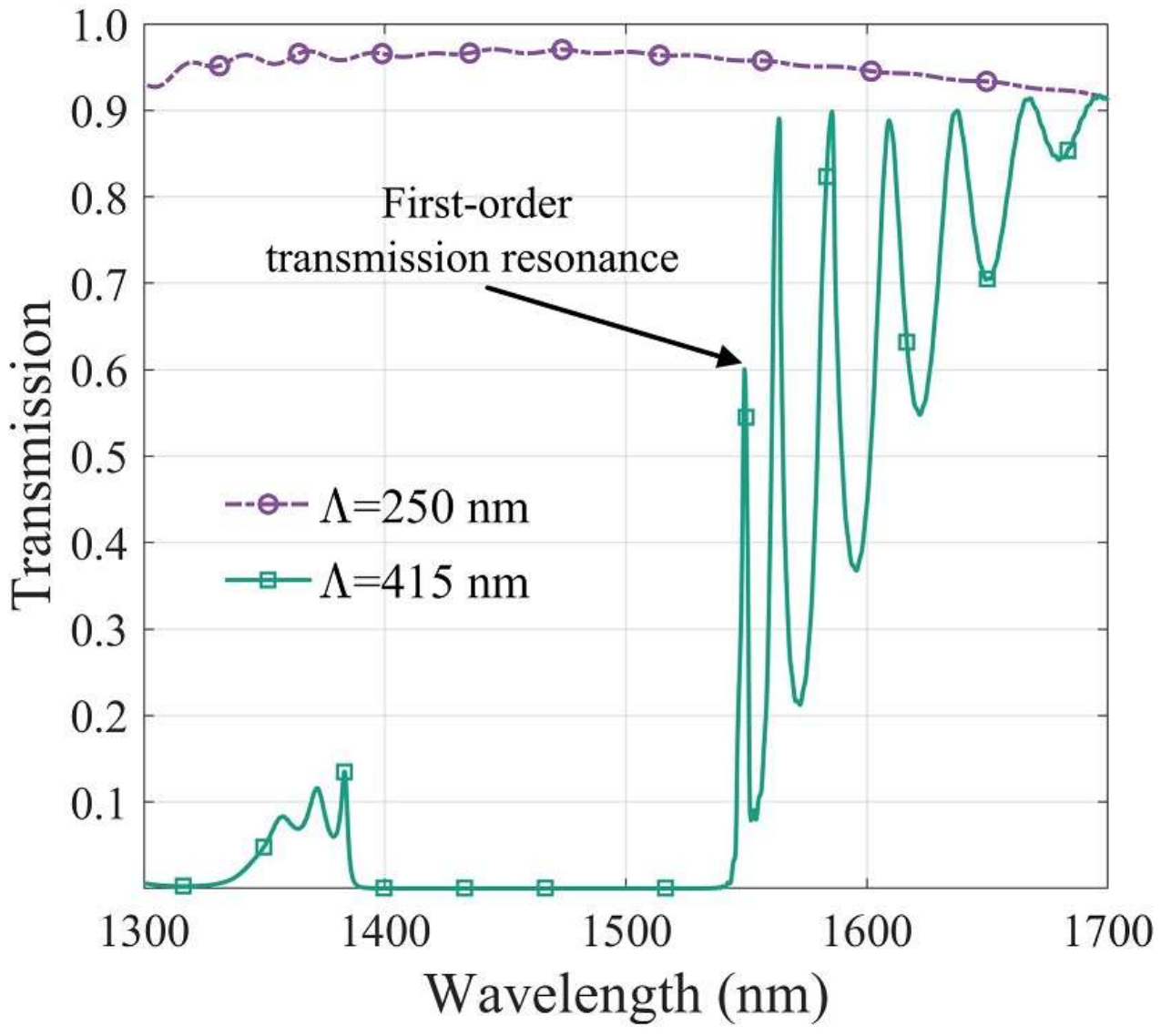

Fig. 2. The transmission spectra of the slot waveguides that operate in different regimes. When the grating pitch is Λ=250 nm, the structure operates in the subwavelength regime. However, for Λ=415 nm, the structure operates in the Bragg reflection regime. The other geometrical parameters are the same for both structures, i.e., $a$=249 nm, $h$=220 nm, $w_{rail}$=400 nm, $w_{slot}$=100 nm, $L_{BG}$=28×Λ and $L_{taper}$=10×Λ.

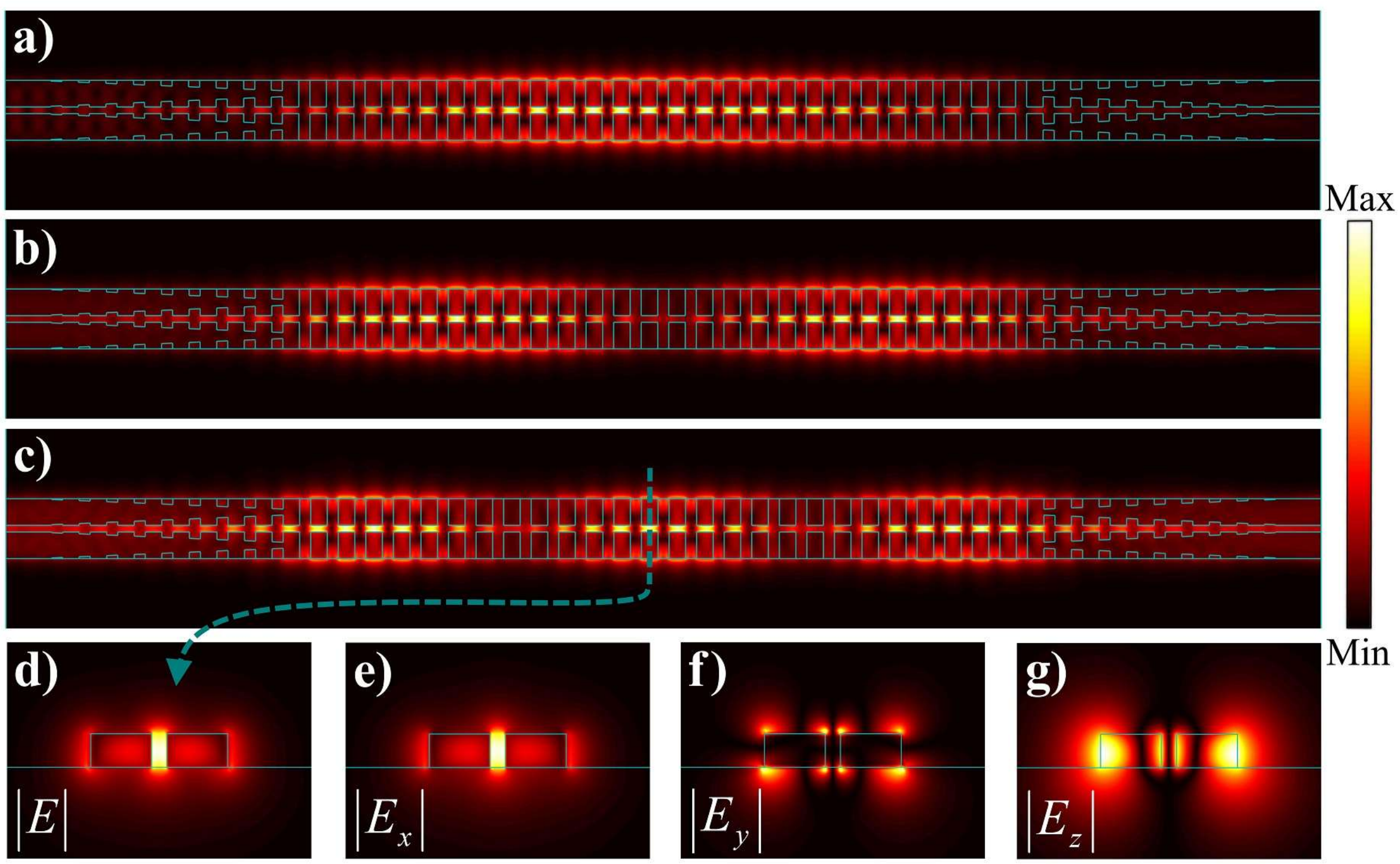

Fig. 3. The electric field intensity of a) the first-order transmission resonance at the wavelength of 1549.29 nm, b) the second-order transmission resonance at the wavelength of 1563.8 nm, and c) the third-order transmission resonance at the wavelength of 1585.7 nm. The electric field distribution of the TE mode and its components at the cross-section of the SWG slot waveguide are displayed in d-g subfigures. The coordinate axes are shown in Fig. 1. The geometrical parameters are Λ=415 nm, $a$=249 nm, $h$=220 nm, $w_{rail}$=400 nm, $w_{slot}$=100 nm, $L_{BG}$=28×Λ and $L_{taper}$=10×Λ.

## 3. Results and discussion

The location of the resonance peaks depends on the geometrical parameters of the SWG slot waveguide. In the following subsections, we investigate the effect of geometrical parameters on the transmission resonances of the proposed structure. In each case, we vary one of the geometrical parameters while the other parameters are fixed. The center of the photonic bandgap of the SWG structure is $\lambda_B=2\Lambda n_{eff}$ where $n_{eff}$ is the effective refractive index of the SWG structure. The bandwidth of the photonic bandgap and the maximum reflection are determined by [27]

$$\Delta\lambda = \frac{\lambda_B^2}{\pi n_g}\sqrt{\kappa^2 + \frac{\pi^2}{L_{BG}{}^2}} \tag{1}$$

$$R_{\max} = \tanh^2(\kappa L_{BG}) \tag{2}$$

where $\kappa$ is the coupling coefficient representing the coupling rate between forward and backward propagating modes of the SWG while $n_g$ is the group index. $\kappa$ and $n_g$ depend largely on SWG geometry [28].

### 3.1 Effect of duty cycle (*D*) on transmission resonances

The duty cycle determines the length of the silicon pillars ($a=D\times\Lambda$). In Fig. 4, the effect of the duty cycle on the transmission resonances is demonstrated. The other geometrical parameters are fixed to Λ=415 nm, $h$=220 nm, $w_{rail}$=400 nm, $w_{slot}$=100 nm, $L_{BG}$=28×Λ=11.62 µm and $L_{taper}$=10×Λ=4.15 µm. As the duty cycle increases the first-order resonance shifts to longer wavelengths and its transmission peak increases. However, the FWHM of the transmission resonance increases. For $D$=0.5, the first-order resonance occurs at the wavelength of 1509.9 nm with the transmission peak of 0.187. The FWHM of this transmission resonance is 2.49 nm. When the duty cycle is set to 0.6, the first-order resonance occurs at the wavelength of 1549.29 nm with the transmission peak of 0.603 while the FWHM is 2.96 nm. As the duty cycle increases to 0.7, the resonance shifts to a longer wavelength of 1584.5 nm with a higher transmission peak of 0.95, however, the FWHM increases to 6.52 nm. The duty cycle determines the effective refractive index of the SWG structure. As $D$ increases, $n_{eff}$ increases, and vice versa. On the other hand, as $n_{eff}$ increases the center of the photonic bandgap and consequently, the transmission resonances shift to lower wavelengths. However, as the duty cycle decreases the transmission peak is reduced considerably. Increasing the duty cycle increases the transmission peak while decreasing the FWHM, therefore, we choose $D=0.6$ with reasonable transmission peak and FWHM.

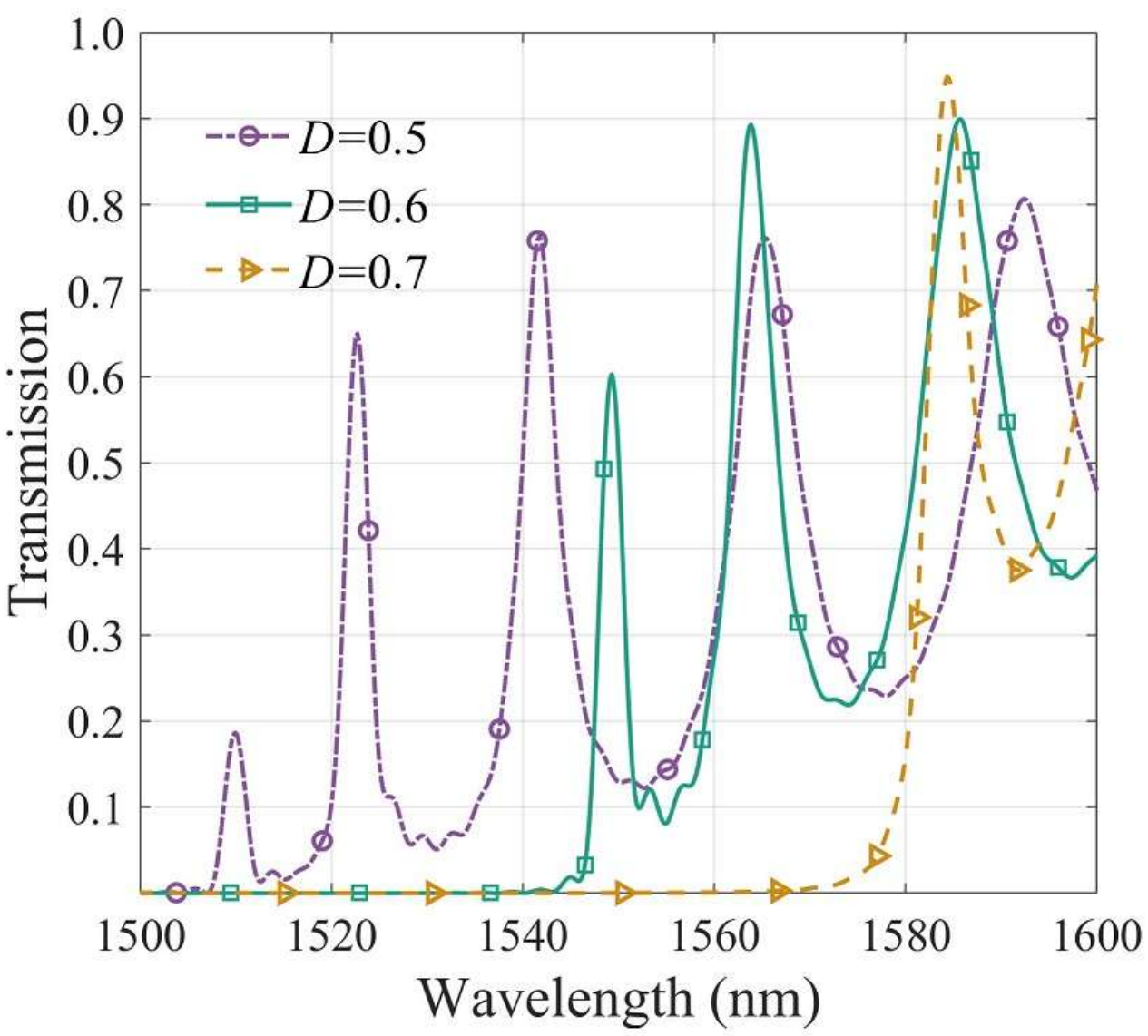

Fig. 4. The effect of the duty cycle on transmission resonances while the other geometrical parameters are set to Λ=415 nm, $h$=220 nm, $w_{rail}$=400 nm, $w_{slot}$=100 nm, $L_{BG}$=11.62 µm and $L_{taper}$=4.15 µm.

### 3.2 Effect of slot gap ($w_{slot}$) on transmission resonances

The light confinement as well as the coupling efficiency depend on the gap between the SWG rails [29]. In Fig. 5, the effect of the slot gap on transmission resonances, while other parameters are fixed to Λ=415 nm, $h$=220 nm, $w_{rail}$=400 nm, $D$=0.6, $L_{BG}$=28×Λ=11.62 µm and $L_{taper}$=10×Λ=4.15 µm, is investigated. As the slot gap increases, the resonance wavelength shifts to shorter wavelengths and the peak transmission increases while the change in FWHM is negligible. For $w_{slot}$=80 nm, the first-order resonance occurs at the wavelength of 1564.18 nm with the transmission peak of 0.407 while the FWHM is 2.82 nm. When the slot gap increases to 100 nm, the first-order resonance occurs at the wavelength of 1549.29 nm with the transmission peak of 0.603 while the FWHM is 2.96 nm. And for $w_{slot}$=120 nm, the first-order resonance wavelength shifts to 1531.1 nm and its transmission peak increases to 0.7, however, its FWHM is 2.95 nm. The propagation loss of a slot waveguide increases as the gap size decreases [30]. On the other hand, as the slot gap increases the mode confinement decreases reducing the sensitivity of the sensor to the changes of the functional material's refractive index. Considering the trade-off between the propagation loss and light confinement, we choose $w_{slot}$=100 nm with lower propagation loss compared to $w_{slot}$=80 nm while offering higher light confinement compared to $w_{slot}$=120 nm.

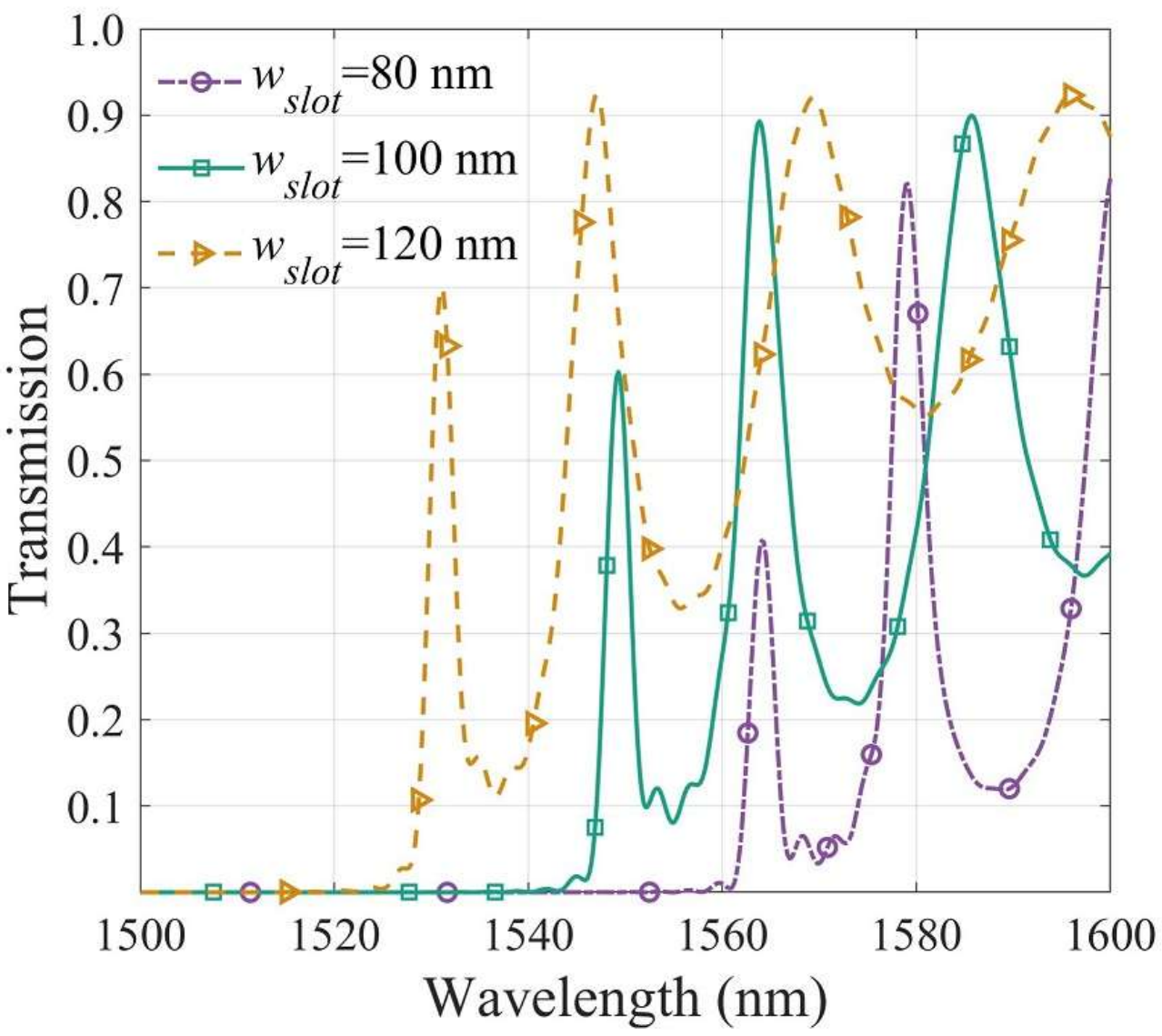

Fig. 5. The effect of the slot gap on transmission resonances while the other geometrical parameters are fixed to Λ=415 nm, $h$=220 nm, $w_{rail}$=400 nm, $D$=0.6, $L_{BG}$=11.62 µm and $L_{taper}$=4.15 µm.

### 3.3 Effect of Bragg grating length ($L_{BG}$) on transmission resonances

At the first-order resonance, constructive interference of the forward and the backward propagating modes in the middle of the SWG structure is obvious in Fig. 3(a). This constructive interference occurs when the propagating wave experiences a round trip of $2\pi m$, where $m$ is an integer, while the phase gathered by the wave is $kL_{BG}$, where $k$ is the wavenumber [31]. Hence, the transmission resonances also depend on the length of the SWG structure. Moreover, as the length of the SWG structure increases the number of transmission resonances increases in the given bandwidth [26]. The length of the Bragg grating determines the transmission peak and the bandwidth of the resonance as illustrated in Fig. 6. For $L_{BG}$=22×Λ=9.13 µm, the first-order resonance is located at 1552.1 nm with a peak of 0.82 while the FWHM is 3.43 nm. As the length of the Bragg grating increases to $L_{BG}$=28×Λ=11.62 µm, the transmission peak at the wavelength of 1549.29 nm decreases to 0.603 while the bandwidth decreases to FWHM=2.96 nm. For $L_{BG}$=34×Λ=14.11 µm, the resonant peak at the wavelength of 1547.4 is 0.38 with the FWHM of 2.76 nm. As the length of the Bragg grating increases the peak of the first-order transmission resonance decreases while its bandwidth gets narrower. For $L_{BG}$=14.11 µm compared to $L_{BG}$=11.62 µm, the decrease in the transmission peak is considerable, however, the FWHM only decreases by 0.2 nm. We choose $L_{BG}$=11.62 µm considering that it has a lower propagation loss compared to $L_{BG}$=14.11 µm while offering narrower FWHM compared to $L_{BG}$=9.13 µm.

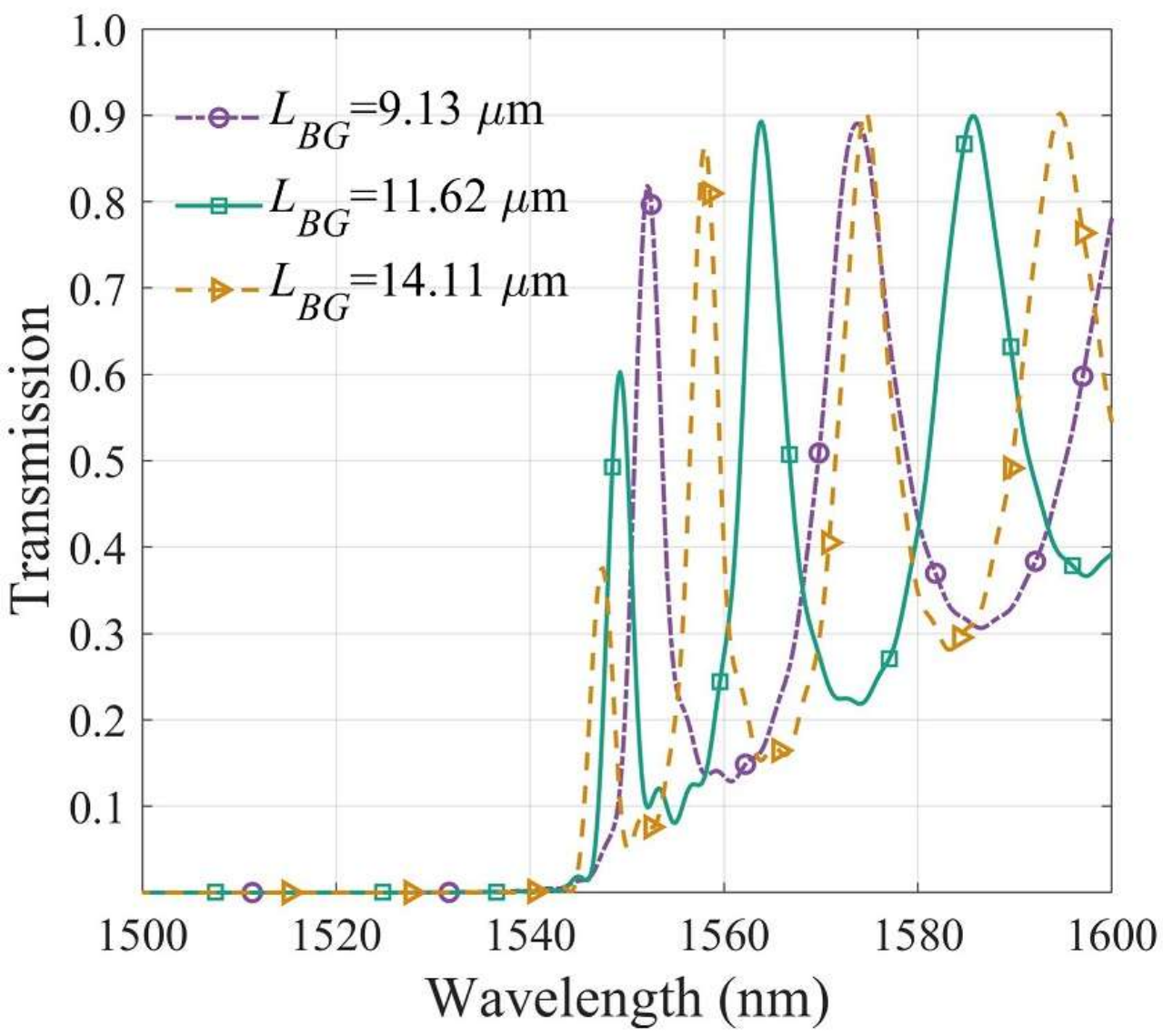

Fig. 6. The effect of Bragg grating length on transmission resonances while the other geometrical parameters are fixed to Λ=415 nm, $h$=220 nm, $w_{rail}$=400 nm, $w_{slot}$=100 nm, $D$=0.6 and $L_{taper}$=4.15 µm.

### 3.4 Effect of grating pitch (Λ) on transmission resonances

The center of the bandgap of the SWG structure is directly proportionate to the grating pitch. Increasing the grating pitch results in the redshift of bandgap's center and consequently, the transmission resonances shift to longer wavelengths. Therefore, we can determine the wavelength of transmission resonances by tuning the grating pitch while keeping the other geometrical parameters fixed. As shown in Fig. 7, the first-order resonance occurs at the wavelength of 1514.9 nm with the peak of 0.57, for the grating pitch of Λ=400 nm. The bandwidth of this resonance is 2.76 nm. For Λ=415 nm, the resonant peak at the wavelength of 1549.29 nm is 0.603 with the FWHM of 2.96 nm. By increasing the grating pitch to Λ=430 nm, the first-order resonance redshifts to 1582.2 nm with the maximum transmission of 0.58 while the FWHM is 3.03 nm. It should be noted that the FWHM increases slightly as the grating pitch of the SWG structure increases. We choose Λ=415 nm to position the first-order transmission resonance in the middle of the C-band.

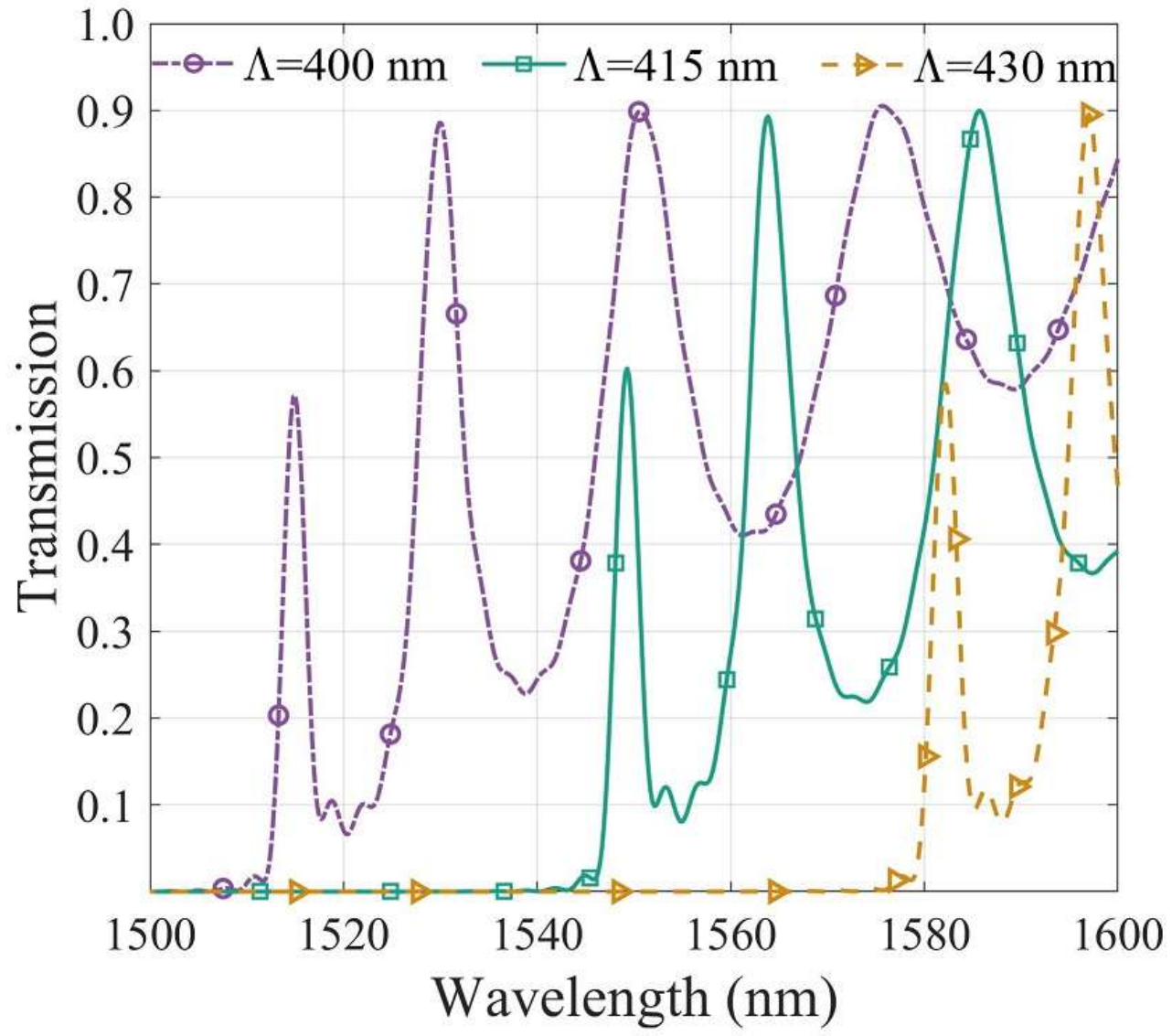

Fig. 7. The effect of grating pitch on transmission resonances while the other geometrical parameters are fixed to $h$=220 nm, $w_{rail}$=400 nm, $w_{slot}$=100 nm, $D$=0.6, $L_{BG}$=26×Λ and $L_{taper}$=10×Λ.

### 3.5 Performance of the designed gas sensor and comparison with previous studies

As discussed in the previous section, the spaces between the silicon pillars as well as the slot are filled with a functional material such as PHMB. The refractive index of PHMB decreases as the concentration of the $CO_2$ gas increases [24]. Here, we examine the performance of the proposed sensor to the refractive index change of PHMB. The refractive index of PHMB decreases due to its exposure to $CO_2$ leading to the shift of the transmission spectrum to shorter wavelengths. The refractive index of PHMB in the absence of $CO_2$ gas is about 1.55 [24]. The decrease of 0.01 in the refractive index of PHMB results in blueshift of the first-order transmission resonance to 1546.51 nm. In reference [24], it is estimated that when the $CO_2$ concentration reaches 215, 262, and 328 ppm, the refractive index of PHMB decreases to 1.54, 1.53, and 1.52, respectively. We evaluate the performance of the proposed structure with other designs based on this estimation. Therefore, the sensitivity of the structure defined as the wavelength change to the concentration of $CO_2$ gas is $S$=2.78 nm/215 ppm=12.9 pm/ppm. And figure-of-merit of the structure is FOM=$S$/FWHM=$4.3\times10^{-3}$ $ppm^{-1}$.

Finally, we compare our design with previous studies. Gas sensors based on the functionalized silicon microring resonator with a 240 nm-thick upper cladding of PHMB have been presented [21, 24]. The resonance wavelength shift of these designs for the $CO_2$ concentration of 262 ppm is less than 1.2 pm which is smaller by a factor of 4000 than our design. The measured sensitivity of the ring resonator presented in [21] is $S$=$2.4\times10^{-3}$ pm/ppm while in our design the calculated sensitivity is $S$=12.9 pm/ppm. In a dual-gas sensor based on microring resonators, the measured sensitivity to the sensor to $CO_2$, coated with a 620 nm-thick PHMB as the functional material, is $S$=$4.83\times10^{-4}$ pm/ppm [22]. The lower sensitivity of this sensor compared to the design of [21] is attributed to the thicker PHMB in this design. In our design, the thickness of the PHMB is limited to 220 nm ensuring that $CO_2$ molecules can easily penetrate deep into the functional material. In another study, a whispering gallery mode microbubble resonator is coated by PHMB [20]. The sensitivity of this sensor to $CO_2$ gas is 0.46 pm/ppm. The superior performance of our designed sensor is due to the fact that the functional material (PHMB) is placed in the core as well as the claddings of the SWG slot waveguide while in previous designs the functional material is utilized as the upper cladding [21, 22, 24]. Since the electric field is largely confined to the core, only a small portion of the field penetrates the upper cladding. This limits the interaction of the field with the functionalized upper cladding reducing the sensitivity of the sensor. On the contrary, we utilize the functional material in the core increasing the proposed structure's sensitivity to the refractive index changes of the functional material.

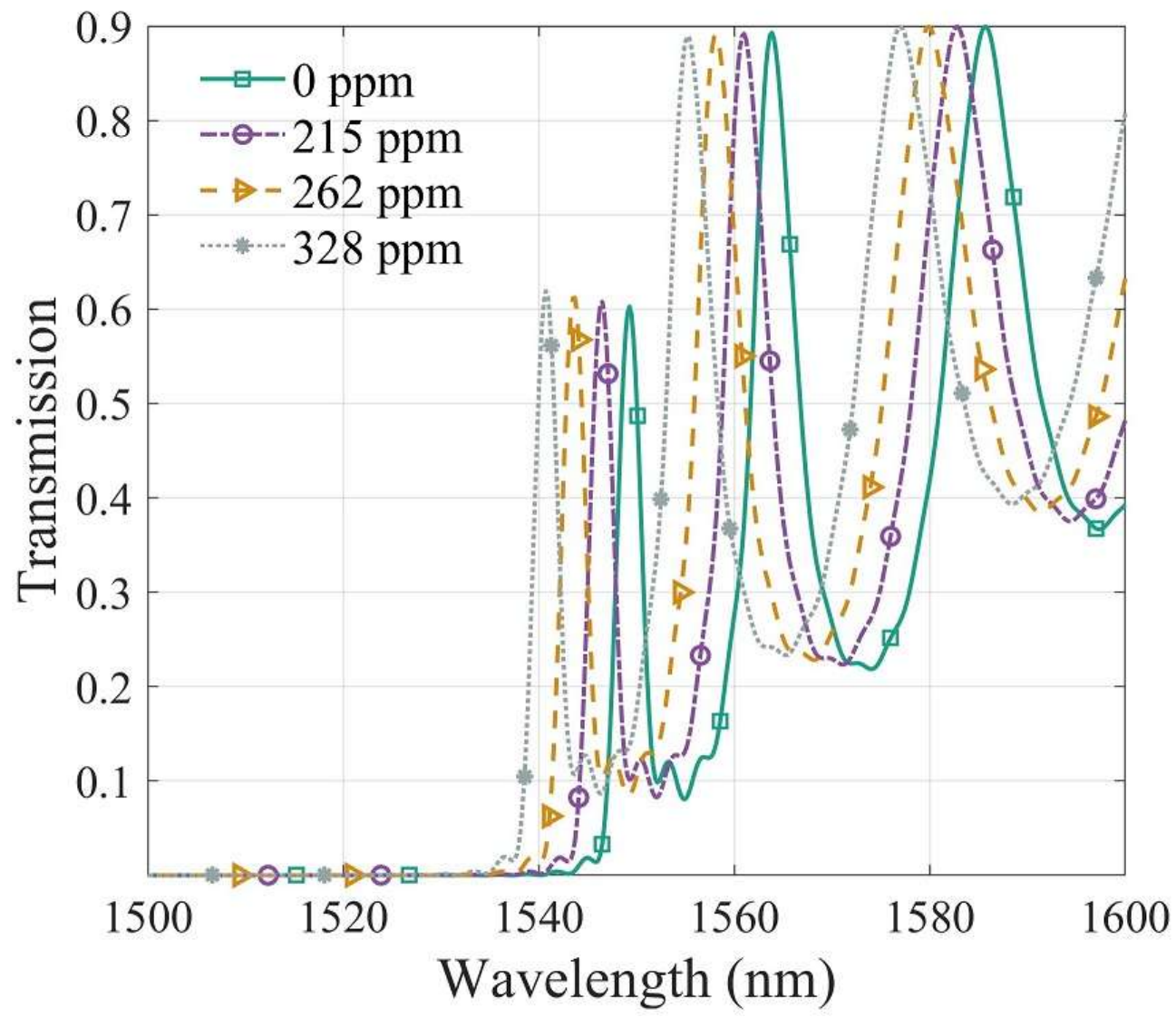

Fig. 8. The resonance wavelength blueshift due to the refractive index change of the functional material (PHMB) in the presence of $CO_2$ gas. The refractive index of PHMB decreases as $CO_2$ molecules penetrate deep into PHMB. The geometrical parameters are $\Lambda$=415 nm, $h$=220 nm, $w_{rail}$=400 nm, $w_{slot}$=100 nm, $D$=0.6, $L_{BG}$=11.62 µm and $L_{taper}$=4.15 µm.

## Conclusion

The sensors based on the slot waveguide typically rely on changes in the modal properties of the waveguide in the presence of the analyte. We present the SWG slot waveguide with transmission resonances. In the proposed structure, the sensitivity to refractive index variation of the functional material is high because the Bloch mode exists between the silicon pillars which is filled with the functional material. On the other hand, the previous designs have lower sensitivity since they rely on the limited interaction of the field and the functional material as the upper cladding. The grating pitch ($\Lambda$) is chosen to have a passband in the C-band while the duty cycle ($D$) and the length of the grating ($L_{BG}$) determine the scattering losses. The calculated sensitivity of the proposed structure due to the refractive index change of the PHMB, as the functional material, in the presence of the $CO_2$ gas is 12.9 pm/ppm. The presented method can be extended to design various sensors with different functional materials.

**Conflict of interest.** The author declares that he has no conflict of interest.

## ReferencesUncategorized References